\documentclass[a4paper]{jpconf}
\usepackage{graphicx}
\newcommand{\mrm}{ }
\newcommand{\ra}{\mbox{$\rightarrow$}}
\begin{document}
\title{Search for the Higgs boson:\\ 
a statistical adventure of exclusion and discovery}

\author{Dezs\H{o} Horv\'ath 
(on behalf of the CMS Collaboration)\footnote{Invited plenary lecture at CCP-2013, the XXV IUPAP Conference on
  Computational Physics, Landau Institute for Theoretical Physics, Moscow, 
21 August 2013.}}

\address{Wigner Research Centre for Physics,
Budapest, Hungary
  \\ and {Atomki, Debrecen, Hungary}}

\ead{horvath.dezso@wigner.mta.hu}

\begin{abstract}
The 40 years old Standard Model, the theory of particle physics, seems to
describe all experimental data very well. All of its elementary particles were
identified and studied apart from the Higgs boson until 2012. For decades many
experiments were built and operated searching for it, and finally, the two
main experiments of the Large Hadron Collider at CERN, CMS and ATLAS, in 2012
observed a new particle with properties close to those predicted for the Higgs
boson. In this talk we describe the search process: the exclusion of the Higgs
boson at LEP, the Large Electron Positron collider, and the observation at LHC
of a new boson with properties close to those predicted for the Higgs boson of
the Standard Model. We try to pay special attention on the statistical methods
used.
\end{abstract}

\section{Introduction: the Standard Model}
The Standard Model, the general theory of particle physics was established
more than 40 years ago. It describes our world as consisting of two kinds of
elementary particles, fermions and bosons, differing by their spin, intrinsic
angular momentum: fermions have half-integer, bosons have integer spins
measured in units of $\hbar$, the reduced Planck constant. The elementary
fermions have three families, each consisting of one pair of quarks and one
pair of leptons. All fermions have antiparticles of opposite charges. The
leptons can propagate freely, but the quarks are confined in hadrons: they can
only exist in bound states of three quarks, baryons (like the proton and
neutron) or those of a quark and an antiquark, mesons (like the pion). Three
antiquarks make antibaryons like the antiproton.

In the Standard Model the three basic particle interactions, the strong interaction
holding the quarks in the nucleons and the nucleons in the atomic nucleus, the
weak interaction, responsible for the decay of nuclei and of the neutron, and
the well known electromagnetic interaction, are all derived from local gauge
symmetries. A gauge symmetry is a freedom to define the coordinate system
measuring the strength of an interaction, the best known example of which is
the freedom to choose the potential zero of an electric field. A local
symmetry is its modified form when the gauge is changing in space-time
according to a known function. The three basic interactions are mediated by
elementary bosons: the strong nuclear force by 8 gluons, the weak interactions
by the three heavy weak bosons and the electromagnetism by the photon. In
order to cancel uncomfortable terms from equations the theory also needs the
existence of an additional scalar boson, a particle with all its quantum
numbers like charges and spin zero.

Local gauge symmetries give correct answers to important questions except the
mass of elementary particles: one has to violate them in order to introduce
their masses. This spontaneous symmetry breaking (SSB) was introduced in
several steps to particle physics and it is now an integral part of the
Standard Model. It is called, somewhat unjustified, also the Higgs mechanism,
although it is the product of several people, so it could also be called 
{\em Higgs--Englert--Brout--Guralnik--Hagen--Kibble} mechanism
\cite{ref:SBB}.

\begin{figure}
\begin{center}
\includegraphics[width=0.5\linewidth]{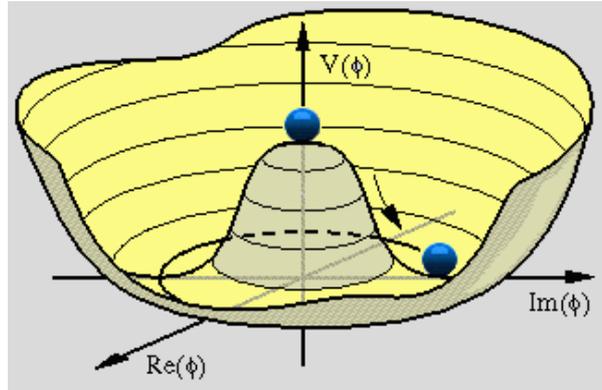}
\end{center}
\caption{\label{fig:ssb}Spontaneous symmetry breaking: the Higgs
  potential. The axial symmetry of the potential is not violated by putting a
  ball on the top at zero, but it will be spontaneously broken when the ball
  rolls down in the valley. However, the coordinate system can always be
  chosen so that the ball were at point $Im(\Phi)=0$. }
\end{figure}

The spontaneous symmetry breaking mechanism consists of adding to vacuum a
potential which breaks its perfect symmetry. This is well illustrated by a
Mexican hat (Fig. \ref{fig:ssb}). Its axial symmetry is not violated by
putting a ball on its top, however, the ball will eventually go down and break
the original symmetry. SSB makes it possible to introduce masses in the
theoretical equations: masses of the heavy weak bosons, W+, W- and Z0
mediating the weak interaction and also masses for the basic fermions, the
quarks and leptons. The masses of the weak bosons are predicted by
the Standard Model, whereas the fermion masses are not, those are free,
adjustable parameters. Note that the masses of our macroscopic world are
mostly due to the energy content of the proton and the neutron and not due to
the SSB mechanism.

\begin{figure}
\begin{center}
\includegraphics[width=0.9\linewidth]{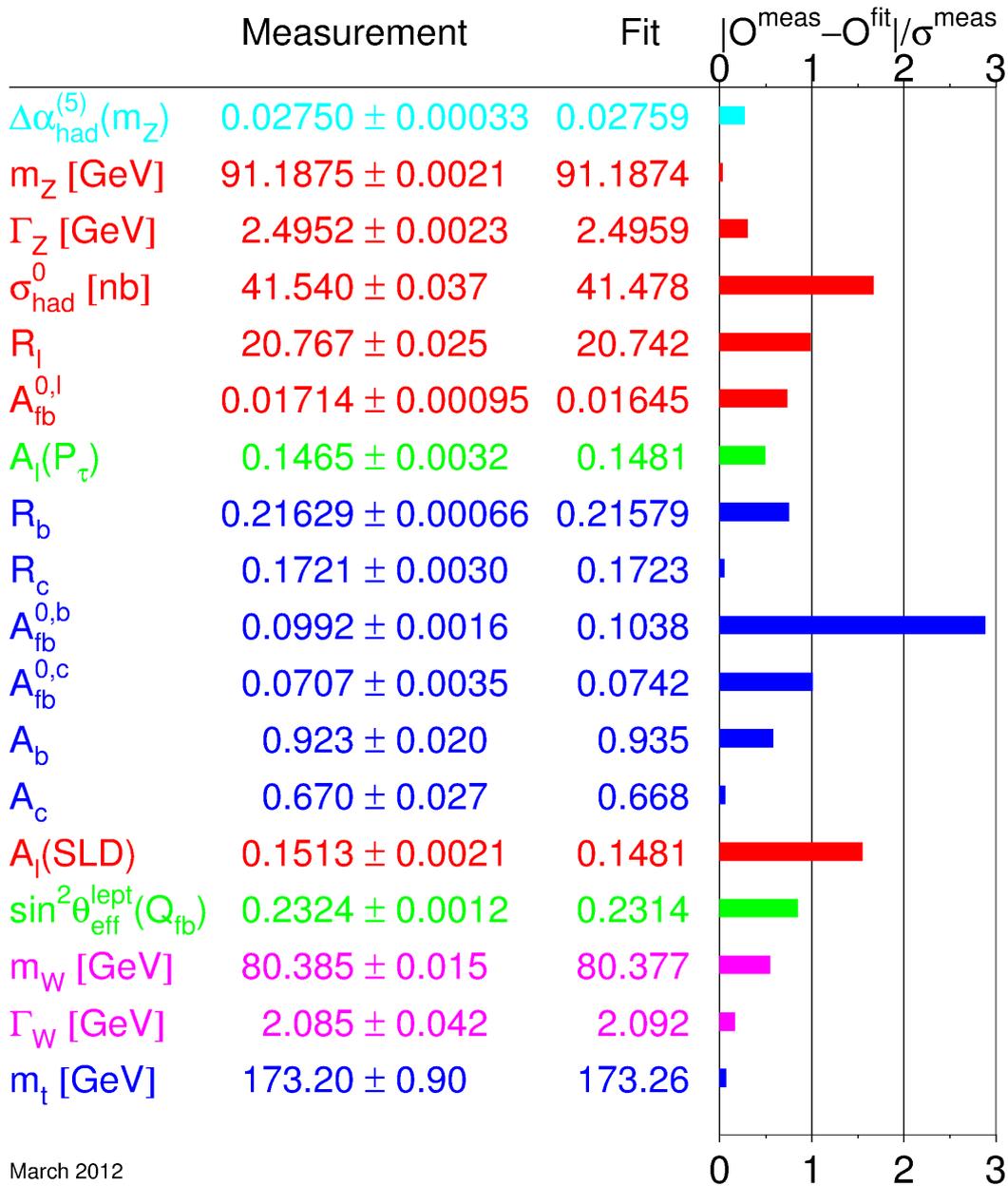}
\end{center}
\caption{\label{fig:show_pull} Parameters of the Standard Model
  \cite{ref:ewwg} as determined by the experiment (2nd column) with
  uncertainties (3rd column), its prediction or fit by the Standard Model (4th
  column) and a bar plot showing the difference between theory and experiment
  divided by the experimental uncertainty. The agreement if purely statistical
  as the difference is in only one case more than 2 uncertainties.}
\end{figure}

Since almost 40 years, more and more precise new data were acquired at the
particle accelerators and all seem to agree very well with the predictions of
the Standard Model. Hundreds of experiments are summarized in
Fig.~\ref{fig:show_pull} according to the LEP Electroweak Working Group
\cite{ref:ewwg}. It shows the 2012 situation of the analysis of electroweak
data: all experimental data and theoretical estimates agree within the
statistical boundaries. The only parameter which deviates at more than 2
uncertainties is the forward-backward asymmetry of 
the decay of the Z boson to two b quarks.

\begin{figure}
\begin{center}
\begin{minipage}{0.80\linewidth}
\includegraphics[width=\linewidth]{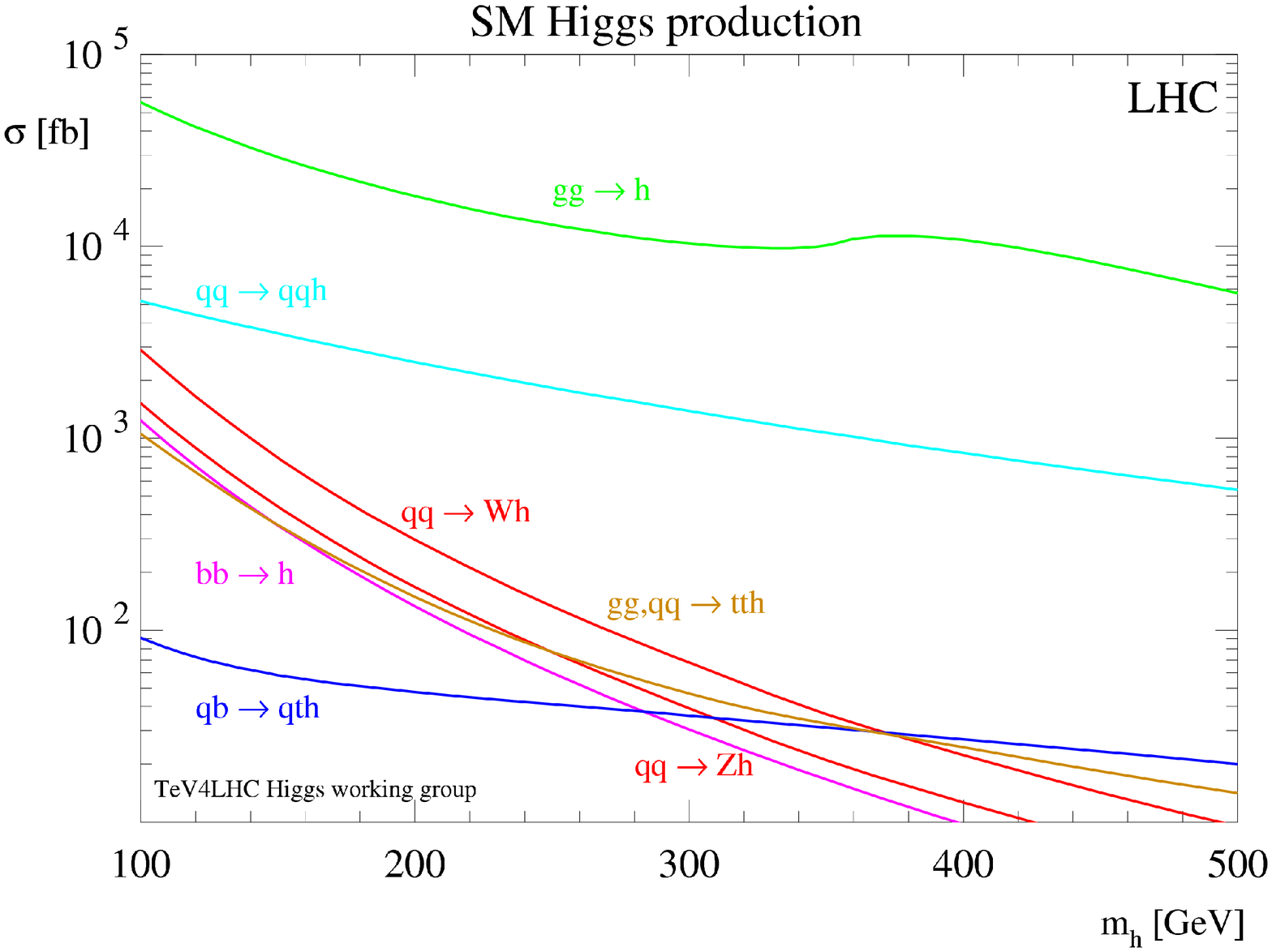}
\end{minipage}
\begin{minipage}{0.15\linewidth}
\begin{center}
\includegraphics[width=\linewidth]{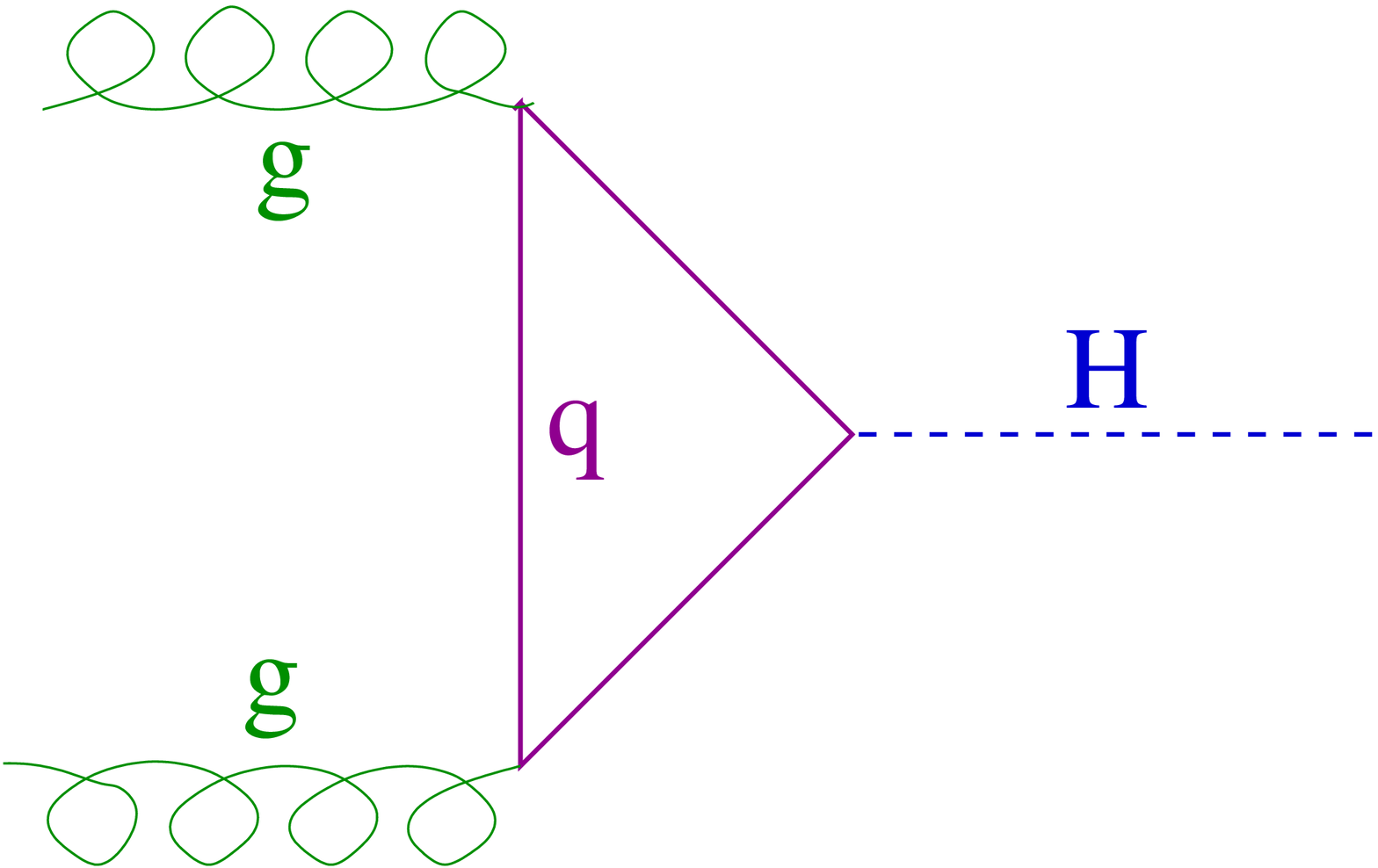}

{\footnotesize
{gluon fusion}

\bigskip\bigskip\bigskip

\includegraphics[width=\linewidth]{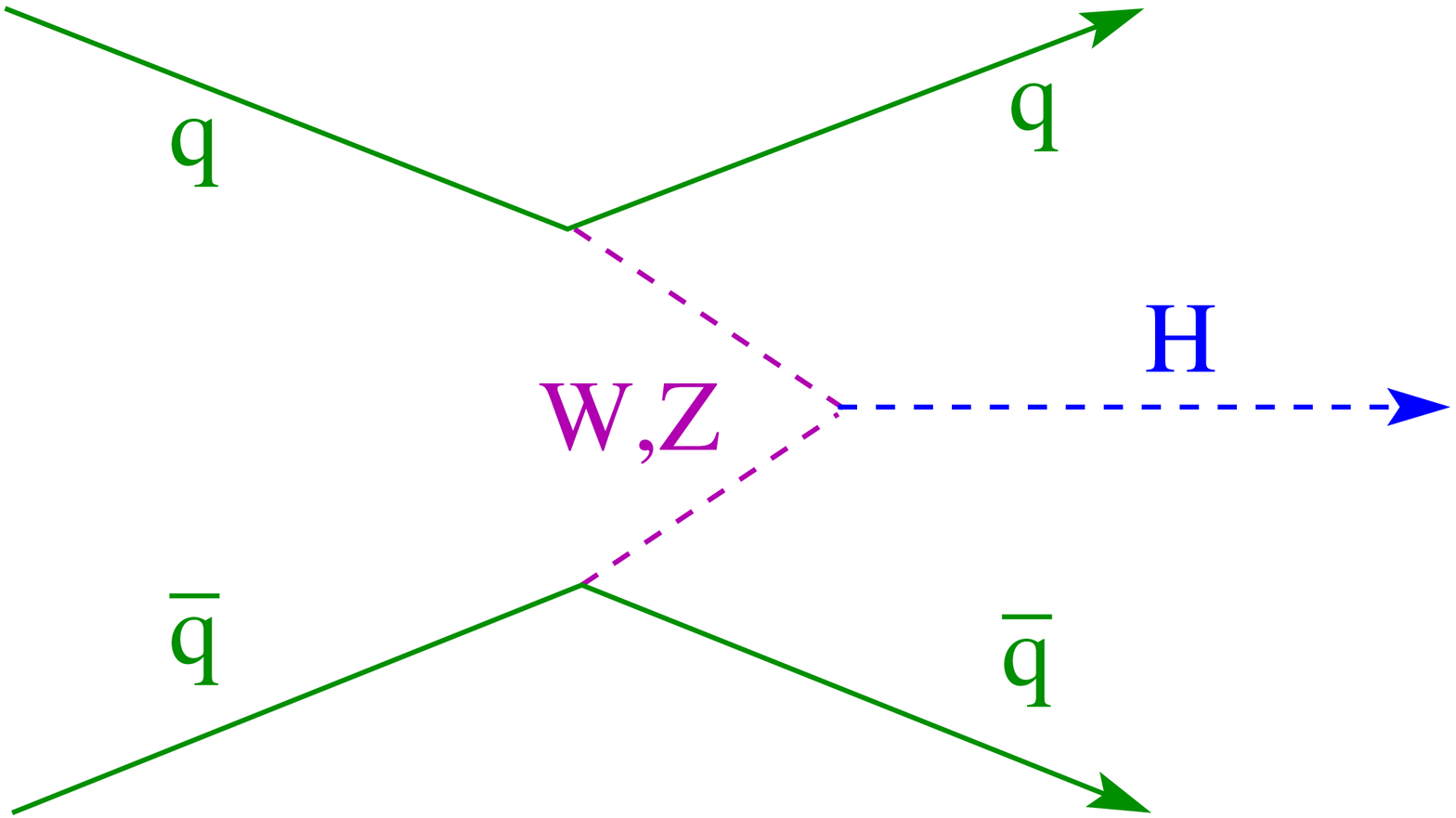}

vector~boson\\ fusion}
\end{center}
\end{minipage}
\end{center}
\caption{\label{fig:SMH-prod}Formation of the SM Higgs boson in p-p collisions
  at LHC.
}
\end{figure}

Figure \ref{fig:SMH-prod} shows the various formation processes of the SM
Higgs boson in p-p collisions at LHC. The dominant reaction is gluon fusion
and vector boson fusion is also significant.

The Higgs boson of the Standard Model is the only scalar particle: all of its
quantum numbers are zero, its only property is mass. Fitting experimental data
predicts that the Higgs mass should be around 100 GeV (between 80 and 160 GeV
within 95\% confidence). All constituents of the Standard Model were
identified and studied experimentally before the launch of the LHC, apart from
the Higgs boson, that is how it became the most wanted particle. As Peter
Higgs himself told \cite{ref:Higgs2002} ``It was in 1972 ... that my life as
a boson really began''.

\section{Statistical Concepts of Particle Physicists}

Those are as different from the {\em official} mathematical statistics as
mechanical engineering from the Lagrangian or Hamiltonian formulation of
theoretical mechanics. At the same time statistics is extremely important for
data analysis in particle physics: every few years international workshops are
organized by particle physicists working at the Large Hadron Collider to
exchange ideas on statistical methods, the last one having been in 2011
\cite{ref:phys-stat-2011}.  In the Appendix of that volume {\em Eilam Gross}
defines the aim of his paper {\em LHC Statistics for Pedestrians}: {\em ''A
  pedestrian's guide $\ldots$ to help the confused physicist to understand the
  jargon and methods used by HEP phystatisticians.  $\ldots$ A phystatistician
  is a physicist who knows his way in statistics and knows how Kendall's
  advanced theory of statistics book looks like.}

Every high-energy collaboration has phystatistician experts and they all have
quite different ideas how to analyze data. In order to avoid confusion, the
large LHC collaborations have Statistics Committees which publish home pages
of recommendations how to do things. The Statistics Committees of both CMS and
ATLAS have several members who published text books on statistics for
physicists and ATLAS and CMS have a joint such committee as well.

As in high energy physics the experimental data are basically event counts,
the basic concepts are Poisson-like. The data follow the Poisson distribution
($n_i$ events in bin $i$): ~ ${\cal P}(n_i|\mu_i) = \frac{\mu_i^{n_i}
  e^{-\mu_i}}{n_i !}$ and the result is usually expressed in terms of the
Poisson likelihood: ${\cal L} = \Pi_i {\cal P}(n_i|\mu_i)$, where the expected
number of events is $\tilde\mu_i = \sum_{j} L \sigma_j \epsilon_{ji}$, $L$ is
the integral luminosity collected,
$\sigma_j$ is the cross section of source $j$ and $\epsilon_{ji}$ is the
efficiency (determined by Monte Carlo simulation) of source $j$ in bin $i$.

Luminosity is the rate of collecting data for colliders, similar to the flux
of fixed-target experiments. It is defined as $L = f n \frac{N_1 N_2}{A}$
where $f$ is the circulation frequency of the colliding beams; $n$ the number
of particle bunches in the ring; $N_1, N_2$ are the numbers of particles in
the two kinds of bunches; $A$ is the spatial overlap of the colliding
bunches. The total number of collisions is characterized by the integrated
luminosity: $\int_{t_1}^{t_2} L dt$ which is usually measured in units of
inverse cross-section, at LHC in $[\mrm{pb^{-1},\; fb^{-1}}]$. The expected
rate of a reaction with cross section $\sigma$ at $\epsilon$ detection
efficiency is $R=\epsilon\sigma L$.

According to the general convention in accelerator experiments a given new
phenomenon is excluded if we can show it not appearing at a $\geq95 \%$
confidence level and observed if it exceeds $>5\sigma$ above background where
now, for a change, $\sigma$ is the experimental uncertainty according to the
best honest guess of the experimentalist.

That $\sigma$ uncertainty has a {\em statistical component} from the number of
observed events and {\em systematic} ones from various sources, like Monte
Carlo statistics and inputs, experimental calibration factors, detection
efficiencies, etc, with the common name {\em nuisance parameters}. To get a
rough estimation of the total error the systematic uncertainties could be
added quadratically to the statistical one: $\sigma= \sqrt{\sigma_\mrm{stat}^2
  + \sigma_\mrm{syst}^2}$, but in practice we derive the final uncertainty via
marginalizing (integrating out) \cite{ref:cous-high} the nuisance parameters
$\Theta$ in likelihood ${\cal L}$ using the related probability distributions
${\cal W}$: ${\cal L}(P;x) = {\cal W}(x|P) = \int {\cal W}(x|P,\Theta) {\cal
  W}(\Theta|P) d\Theta$.

Another important feature of high-energy data analysis is the {\em blind
  analysis} \cite{ref:blind}: ``A blind analysis is a measurement which is
performed without looking at the answer. Blind analyses are the optimal way to
reduce or eliminate experimenter's bias, the unintended biasing of a result in
a particular direction.'' It came from medical research and the idea is to
optimize, prove and publish your analysis technique using simulations and
earlier data only before touching new data in the critical region. For
instance, in Spring and early Summer, the 2012 CMS data were blinded in the
$110< M_H < 140$ GeV (where $M_H$ is the simulated Higgs mass) because of the
$3\sigma$ excess observed in 2011. The same procedure was used again in Autumn
2012. The methods had to be fixed and approved by the collaboration before
simultaneous {\em unblinding} for all analysis channels.

\section{Search for the Higgs boson}

What we usually try to observe is a resonance. For a particle with lifetime
$\tau=\Gamma^{-1}$ and decay rate $\Gamma$ the event rate against the
invariant mass of the decay products is $|\chi(E)|^2
=\frac{1}{(E-M)^2+\Gamma^2/4}$, i.e. a Lorentz curve (Breit-Wigner
resonance). It shows a peak at the $M$ invariant mass of the decaying system
with a full width at half maximum $\Gamma$. We claim the discovery of a 
new particle if we see a resonance at the invariant mass of the particle in
all expected decay channels, by all related experiments. 

The search involves several consecutive steps.
\begin{itemize}
\item Compose a complete {\em Standard Model background} using Monte Carlo
  simulation taking into account all types of possible events normalized to
  their cross-sections.
\item Compose {\em Higgs signals}, simulations of all possible production and
  decay processes with all possible Higgs-boson masses.
\item Put all these through the {\em detector simulation} to get events
  analogous to the expected measured ones.
\item {\em Optimize the event selection} via reducing the  $B$  background and 
enhancing the $S$ signal via maximizing e.g.\ 
$N_S/\sqrt{N_B}$ or $2\cdot(\sqrt{N_S+N_B}-\sqrt{N_B})$ \cite{ref:BitKras}
\item {\em Check the background}, i.e.\ the description of data by the
  simulation for the given luminosity: the simulation should reproduce the
  observed background distributions in all details. For instance, you can
  check the background of the decay of a neutral particle to charged leptons
  by selecting lepton pairs of identical charges.
\item {\em Check the signal:} does it agree with the expectation by the
  theoretical model?
\end{itemize}

Once you are happy with the simulations and the event selection, you must
chose a test statistic. That could be any kind of probability variable
characteristic of the given phenomenon: probabilities for having background
only, signal or combinations. One of the favorite is the $Q$ likelihood ratio
of signal + background over background: $Q = {\cal L}_{s+b}/{\cal L}_{b}$. As
you see, although our basic approach is definitely frequentist there is a
certain Bayesian influence as well.  What most frequently plotted is 

$-2 \ln
Q(m_H) = 2 \sum_{k=1}^{N_\mrm{ch}} \left[s_k(m_H) - \sum_{j=1}^{n_k}
  \ln\left(1+\frac{s_k(m_H)S_k(x_{jk};m_H)}{b_kB_k(x_{jk})}\right) \right]$

Here the variables are the following:
\begin{itemize}
\item $n_k$: events observed in channel $k$, ~~ $k=1\ldots N_\mrm{ch}$.
\item $s_k(m_H)$ and $b_k$: signal and background events in channel $k$ for
  Higgs mass $m_H$.
\item $S_k(x_{jk};m_H)$ and $B_k(x_{jk})$: probability distributions for
  events for Higgs mass $m_H$ at test point $x_{jk}$.
\item $x_{jk}$: position of event $j$ of channel $k$ on the plane of its
  reconstructed Higgs mass and cumulative testing variable constructed of
  various special features of the event like b-tagging, signal likelihood,
  neural network output, etc.
\end{itemize}

\begin{figure}
\begin{center}
\includegraphics[width=0.6\textwidth]{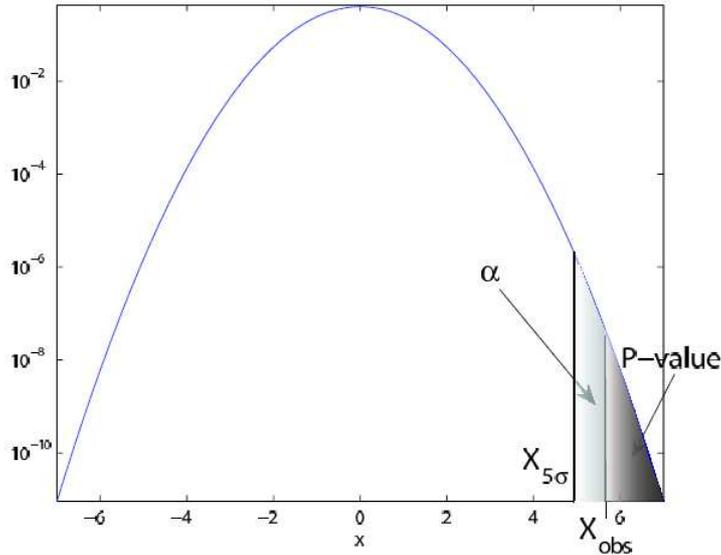}
\caption{\label{fig:p-value_def}The definition of the $p$-value: how much is
  the observed excess above background as compared to the $\sigma$
  uncertainty. We claim a discovery if that excess is above $5\sigma$.}
\end{center}
\end{figure}

Several other testing variables can be constructed on the same basis, the most
frequently used ones are probabilities of NOT having the expected signal on
the basis of the expected background and the collected data:
\begin{itemize}
\item $CL_b$, the signal confidence level assuming background only, i.e. the
  complete absence of the signal, or
\item The so-called {\em $p-$value}: the probability of obtaining a test
  statistic at least as extreme as the one that was actually observed,
  assuming that the null hypothesis is true
  (Fig.~\ref{fig:p-value_def}). Translated to our language that means the
  probability that random fluctuation of the measured background could give
  the observed excess.
\end{itemize}

\section{Exclusion at LEP}

Although the four large experiments at the Large Electron Positron (LEP)
collider saw no new physics, no deviation from the Standard Model, LEP
provided an incredible amount of very precise measurements, some of which are
presented in Fig.~\ref{fig:show_pull}. In its last two years of working, LEP
was mostly devoted to the search for the Higgs boson, collecting more
luminosity at higher energies than in the previous 10 years together.

\begin{figure}
\begin{center}
\includegraphics[width=0.6\linewidth]{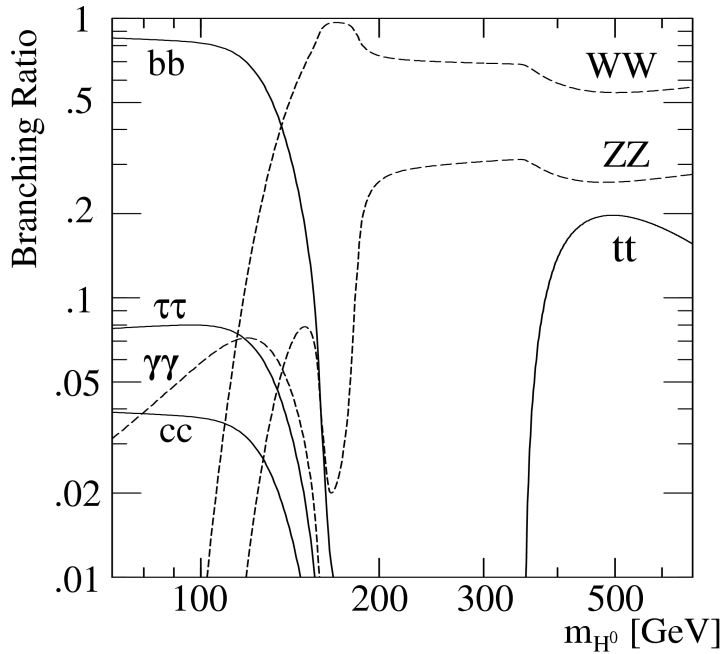}
\caption{\label{fig:SMH_br1}The various decay channels of the Higgs boson
  according to the Standard Model. Below 120~GeV the $\mathrm{H
    \ra\ b\overline{b}}$ decay channel dominates.}
\end{center}
\end{figure}

At LEP the dominant formation process is {\em Higgs-strahlung} 
$\mrm{e^-e^+\ra ZH}$ (the name comes from the funny {\em English} word
Bremsstrahlung) and the dominant Higgs decay is to 2 b-quarks. The various 
channels are different only due to the various decay processes of the
accompanying Z boson.

LEP had 4 large experiments in the 4 interaction points of the
electron-positron collider, ALEPH, DELPHI, L3 and OPAL (the present author was
in OPAL). The structure of the large high-energy detectors are very similar,
consisting of onion-like layers. A sensitive pixel detector right around the
beam pipe, a tracking system of multiwire chambers or semiconductor detectors
of minimal weight material following the tracks of the particles in the
magnetic field of the detector, then an electromagnetic calorimeter, something
heavy absorbing all electrons and photons, outside of that an even heavier
hadron calorimeter, absorbing the pions, protons, neutrons, etc., and finally,
muon chambers, identifying the path of muons leaving the system. All detectors
have huge magnets encompassing as much as possible of the detector parts.

\begin{figure}[t]
\begin{center}
\includegraphics[width=\linewidth]{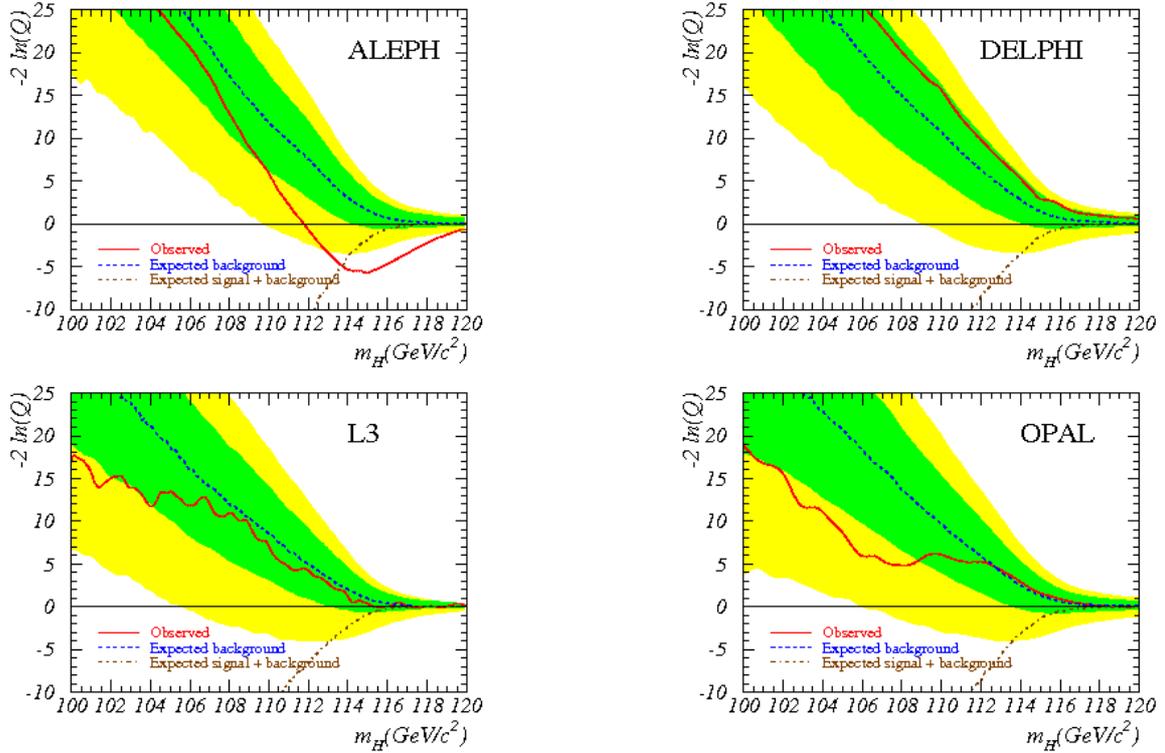}
\end{center}
\caption{\label{fig:adlo_2lnq}Exclusion of the Higgs boson at LEP. The test
  statistic, $-2\ln{Q}$ shows a significant signal for ALEPH and nothing for
  the other 3 LEP experiments at equivalent statistical and experimental
  circumstances. The observed signal of ALEPH by far exceeds the expectations
  of the Standard Model.
}
\end{figure}

Statistics played a rough joke at LEP: one of the experiments, ALEPH, saw in
one of the possible Higgs decay channels a very significant signal
corresponding to a Higgs boson of a mass of 115~GeV/$c^2$, while the rest of
LEP have not seen anything (Fig.~\ref{fig:adlo_2lnq},
\cite{ref:lephiggs2003}). ALEPH saw the excess in the 4-jet events only, in
those events where the Higgs boson decays to a pair of b quarks and the
accompanying Z boson also decays to a quark pair. The b quark is identified
by its long lifetime leading to a secondary decay vertex in the event.
Another strange thing was that the Higgs signal seen by ALEPH by far exceeded
the expectations of the Standard Model. Also, the observed Higgs mass was
critical as it coincided with the average kinematic limit of LEP: in 2000
the average collision energy of LEP was about 206 GeV and the observed
resonance was found at 115~GeV/$c^2$, the difference is very close to the mass
of the Z boson, 91~GeV/$c^2$. 

\begin{figure}[t]
\begin{center}
\includegraphics[width=0.4\linewidth]{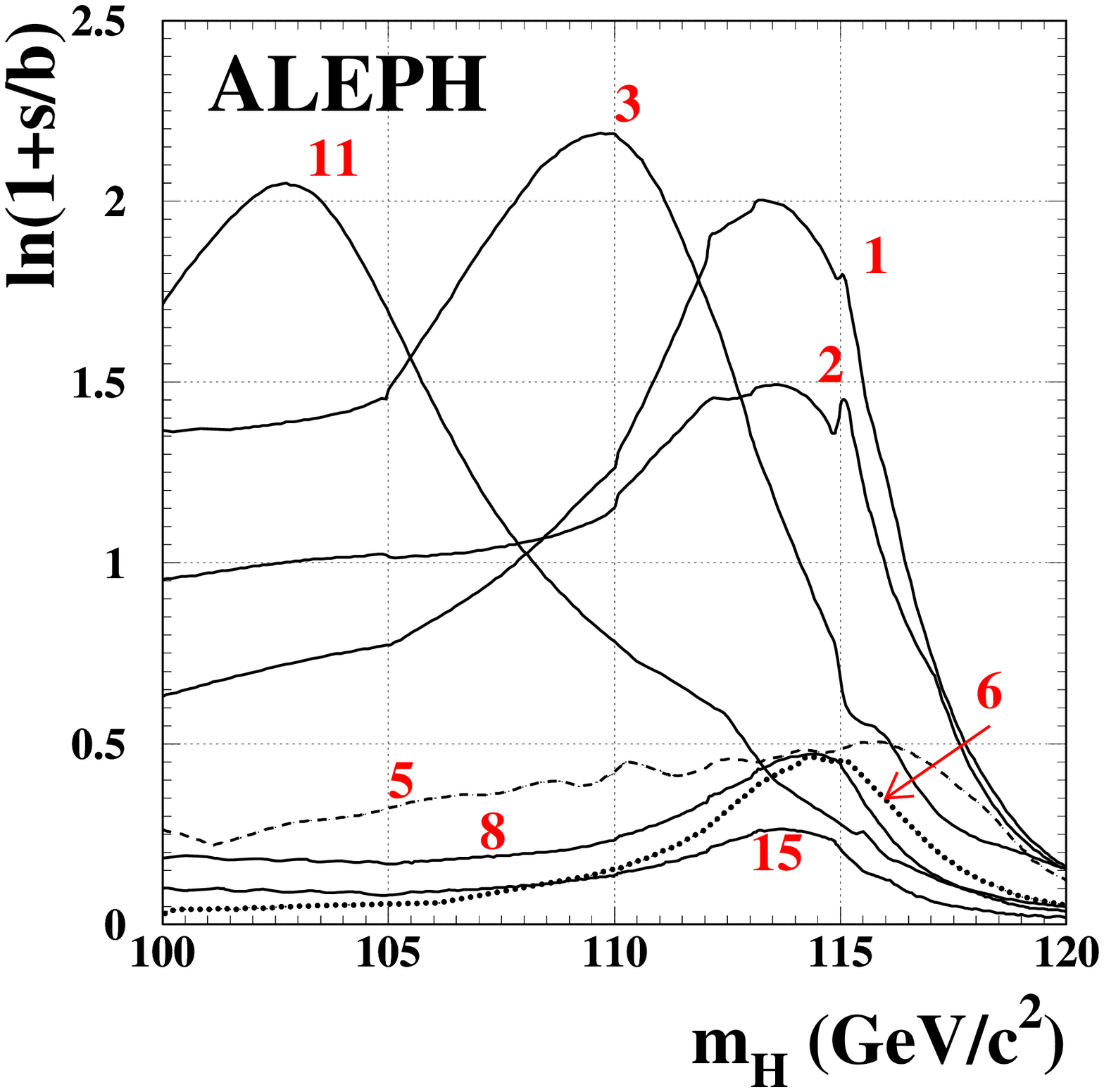} 
\includegraphics[width=0.4\linewidth]{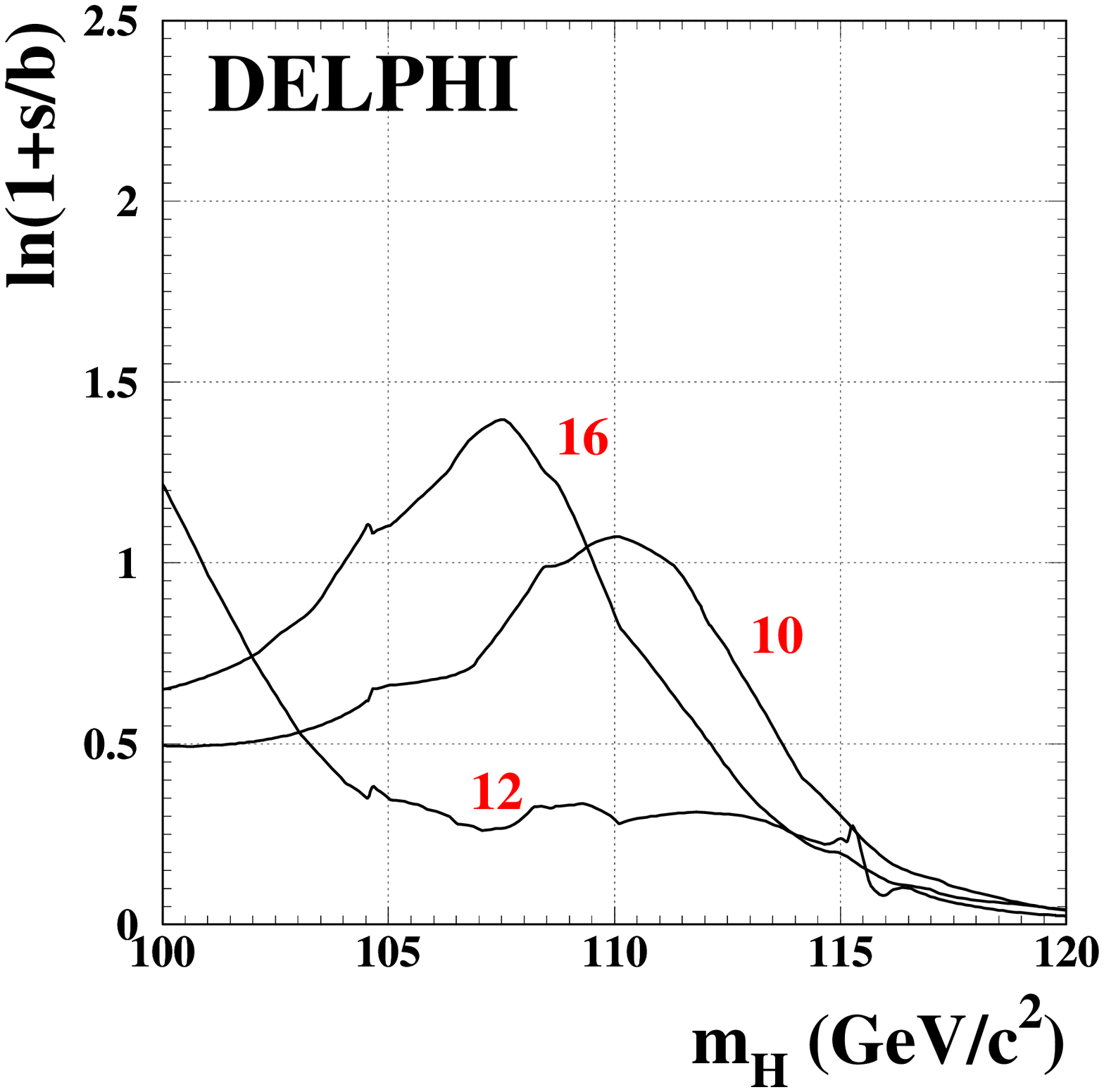}

\includegraphics[width=0.4\linewidth]{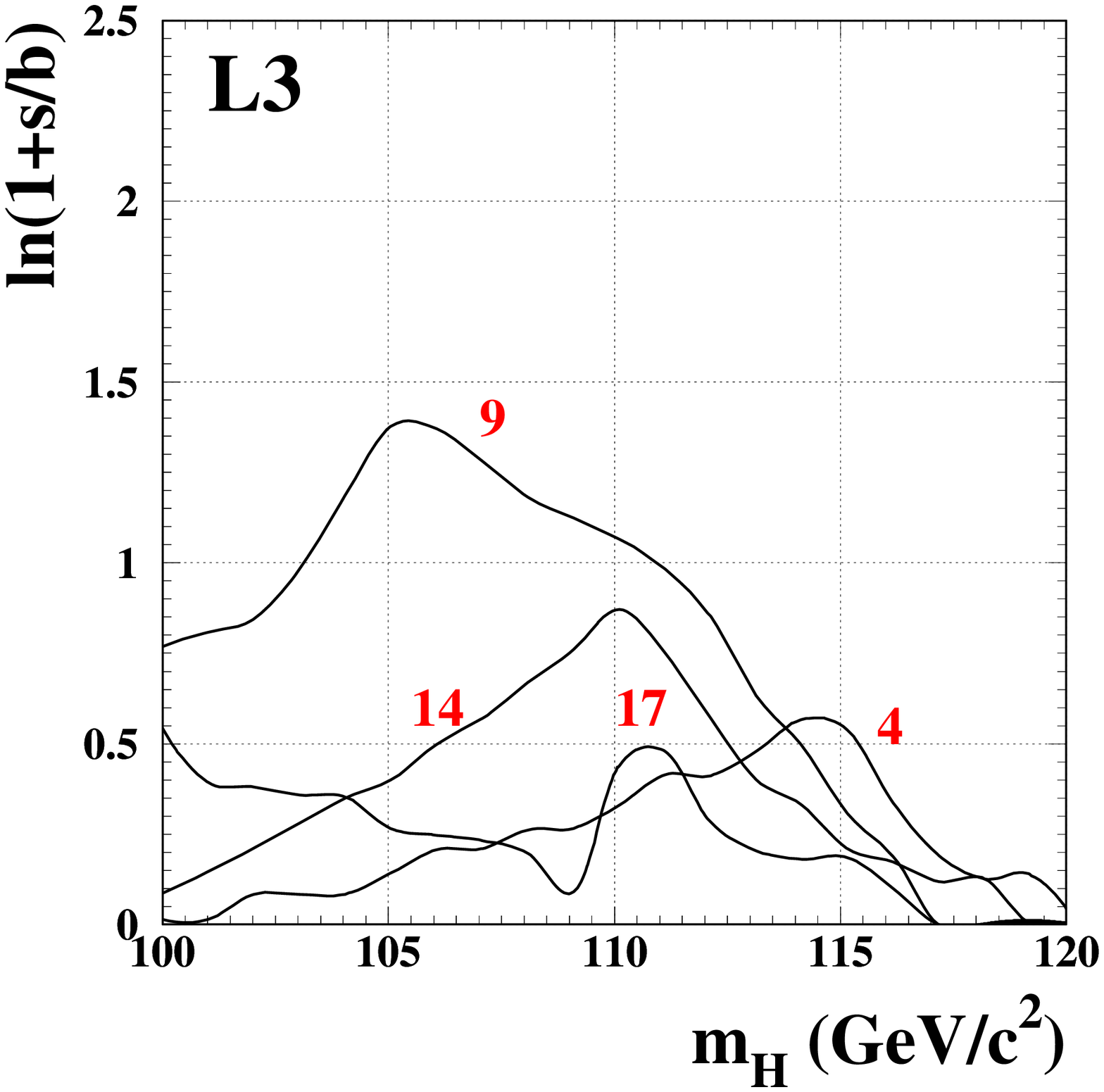} 
\includegraphics[width=0.4\linewidth]{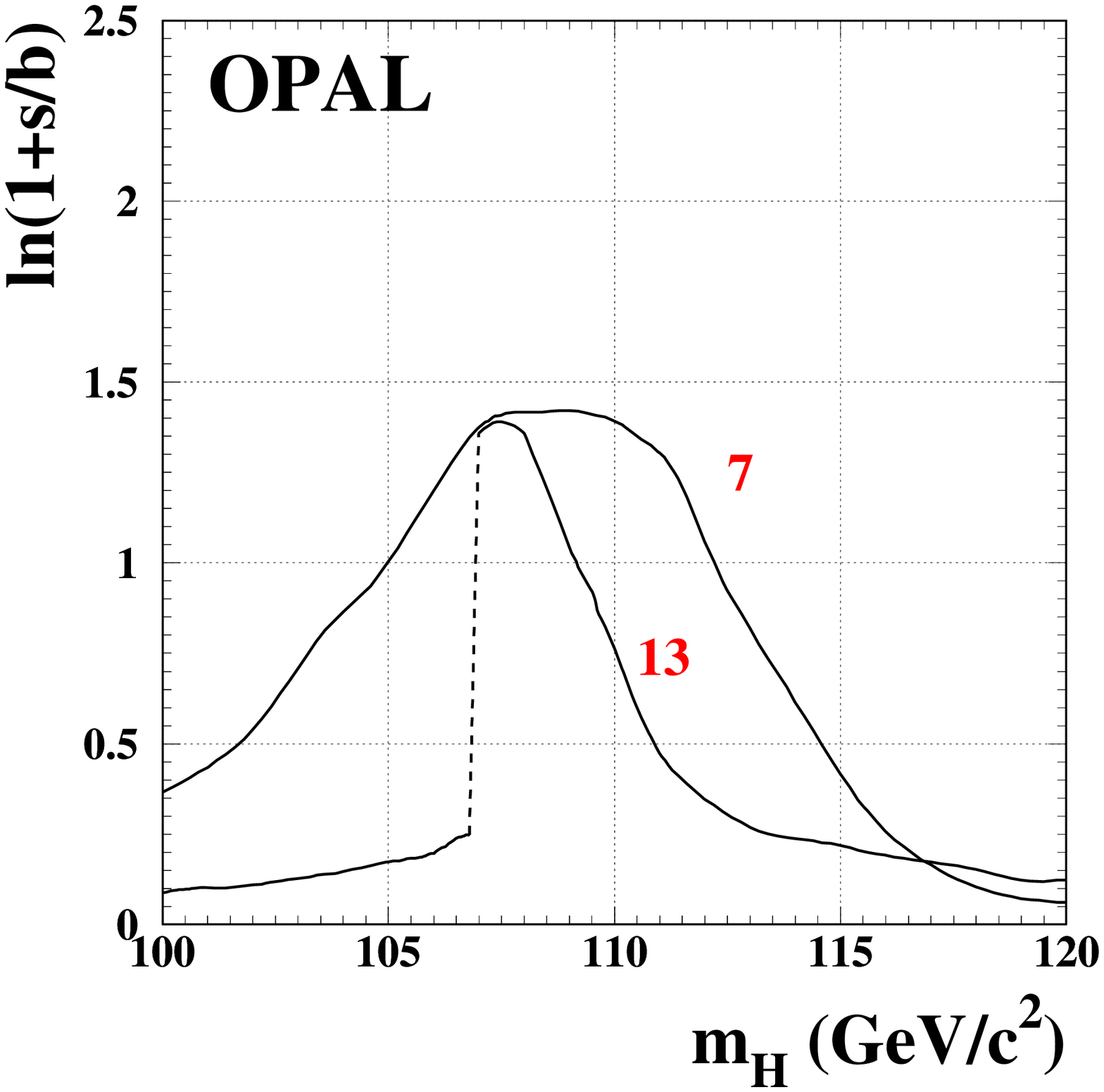}
\end{center}
\caption{\label{fig:spaghetti}{\em Spaghetti diagrams} of 17 selected events
  for the 4 LEP experiments: signal weights against the simulated Higgs mass
  \cite{ref:lephiggs2003}. The ALEPH events crowd around 115~GeV/$c^2$ whereas
  for the other 3 experiments there are less of them with a rather random mass
  distribution.}
\end{figure}

A quite interesting feature of data analysis was the plotting of {\em
  spaghetti diagrams}. Those are signal weight distributions of each selected
event as a function of the assumed Higgs mass. Fig.~\ref{fig:spaghetti} shows
the weight distributions of 17 selected Higgs-like candidate events observed
by the 4 LEP experiments \cite{ref:lephiggs2003}. The ALEPH events crowd around
115~GeV/$c^2$ whereas for the other 3 experiments there are less of them with
a rather random mass distribution. This caused quite an excitement at LEP:
many physicists signed the petition to the Director General of CERN to extend
the life of LEP by another year, but that was refused: the simulated
projections were not very promising for a discovery of the SM Higgs boson (the
effect seen by ALEPH only was far too large, much higher than the prediction
of the Standard Model), and the contractors for building LHC were already
prepared to start.

\section{Observation at LHC}

Just like LEP had, the Large Hadron Collider has also 4 interaction points
with a major experiment (and possibly smaller ones) in each. The two largest
ones, ATLAS \cite{ref:ATLAS} and CMS \cite{ref:CMS} were designed with the
main aim of discovering the Higgs boson, ALICE \cite{ref:ALICE} is specialized
on heavy ion collisions and LHCb \cite{ref:LHCb} on studying rare processes
involving b quarks. The author belongs to CMS, so most of the results we
mention are due to CMS, but all will be compared to those of ATLAS pointing
out the similarities and the (very few and not significant) differences.
These collaborations are huge. According to the official statistics in 2012
CMS had 3275 physicists (incl. 1535 students) and 790 engineers and
technicians from 179 institutions of 41 countries. The largest participant
country was the USA, then Italy, Germany and Russia.

The design of LHC and its experiments started well before the actual start
of LEP, which means that the construction of the LHC detectors took two 
decades of hard work before the actual data acquisition started. Its first
two years LHC devoted to development rather than data taking, that really
started in 2011 only. 

Even before LHC started the parameter fitting of the Standard Model pointed
toward a light Higgs boson, with a mass around 100~GeV/$c^2$. As LEP excluded
the Higgs boson below 114~GeV/$c^2$ the LHC experiments had to be prepared for
detecting the Higgs boson in the most complicated mass region, around
120~GeV/$c^2$, with several competing decay channels. It was shown very early
that the best channel to observe a light Higgs boson at LHC should be the
decay to 2 photons because of the very high hadron background. Thus both large
experiments, CMS and ATLAS designed their electromagnetic calorimeters with
this in mind. The CMS one consists of 75,848 PbWO$_4$ single crystal
scintillators, whereas the electromagnetic calorimeter ATLAS is a sampling one
based on liquid argon shower detectors.

By the beginning of 2012, when all 2011 data were analyzed, the possible mass
of the SM Higgs boson was already confined to the region of 114 $< M_\mrm{H}<$
127 GeV/$c^2$ by CMS~\cite{ref:CMS_H-comb_2011} (with very similar results
from ATLAS). In that region $2\ldots 3 \sigma$ excesses were found at
$\sim$125 GeV/$c^2$ in the two main decay channels, H~\ra$\gamma\gamma$ and
H~\ra~ZZ. After reanalyzing their data the Tevatron experiments, CDF and D0
also found an excess at this mass (after the LHC started the Tevatron
accelerator of Fermilab was stopped). It seemed more and more probable that
the Higgs boson will be observed at LHC in 2012, it was even decided by the
CERN administration to extend the data taking scheduled for 2012 before the
long shutdown for accelerator development if necessary for the discovery.

July 4th, the beginning of the large annual high-energy physics congress in
Melbourne, the spokespersons of ATLAS and CMS gave talks from CERN (in internet
connection to the whole word, including, of course, the main auditorium of the
Australian conference) on Higgs search. They announced that at LHC collision
energies 7 and 8 TeV, in two decay channels H~\ra~$\gamma\gamma$ and
H~\ra~ZZ~\ra~$\ell^+\ell^-\ell^+\ell^-$, at an invariant mass of {$m\approx
  126$ GeV} a new boson is seen at a convincing statistical significance of
$5\sigma$ confidence level each with properties corresponding to those of the
Standard Model Higgs boson. The fact that the new particle could decay to two
photons or Z bosons, confined its spin to an even integer, i.e. a boson of
$S=0$ or $S=2$. Of course, as the data analysis was optimized to find the SM
Higgs, it was very unlikely to find something very different. Nevertheless,
the two experiments emphasized that it has to be studied, whether or not its
spin is really zero with a + parity (the pseudo-scalar mesons have spin 0 with
negative parity), and that its decay probabilities to various final states
follow the predictions of the Standard Model.

\section{Reactions of the Media}

 The saying that {\em three people can keep something secret only if two of
   them are dead} is attributed to Benjamin Franklin. As any result of a
 collaboration has to be approved by all members before it is made public, the
 more than 6000 participants of two large experiments knew well in advance the
 developing result. Thus two days before the 4th July announcement, {\em
   Nature Online} already reported the result~\cite{ref:Nature2July}.  Of
 course, the fact that CERN invited all leading scientists of the field
 including those who developed spontaneous symmetry breaking for the Standard
 Model also helped people to guess that something dramatic will be announced.

 CERN produced some figures concerning the media echo of the day: 55 media
 organizations were represented at the talks of 4 July, the talks were
 broadcasted via close to half a million internet connections (many of them
 conference rooms in partner institutions, e.g. three in Hungary with quite an
audience in each), 1034 TV stations devoted 5016 news broadcasts to the event 
for more than a billion ($10^9$) people. Many-many news articles and even more
blogs and talks discussed the conditions and importance of the discovery.

\section{The observations}

\begin{figure}[t]
\begin{center}
\includegraphics[width=0.7\linewidth]{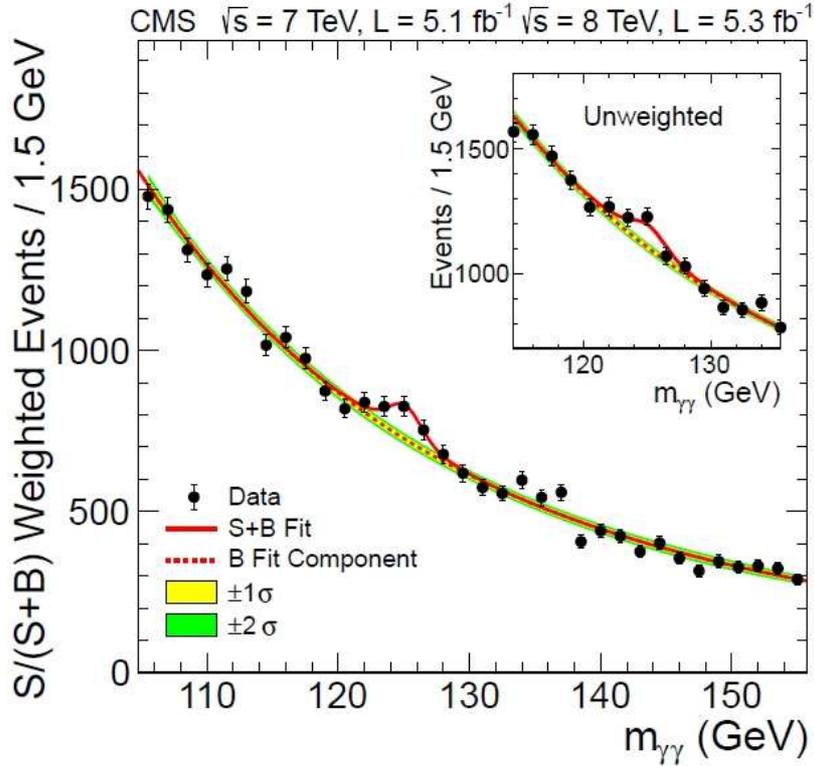}
\end{center}
\caption{\label{fig:cms_h2gg}Observation of the Higgs-like boson by CMS
  \cite{ref:cms_higgs_20120817} in the $\gamma\gamma$ invariant mass
  distribution at 125 GeV/$c^2$. The amplitude of the observed signal is close
  to the expectations of the Standard Model.}
\end{figure}

\begin{figure}[t]
\begin{center}
\includegraphics[width=0.7\linewidth]{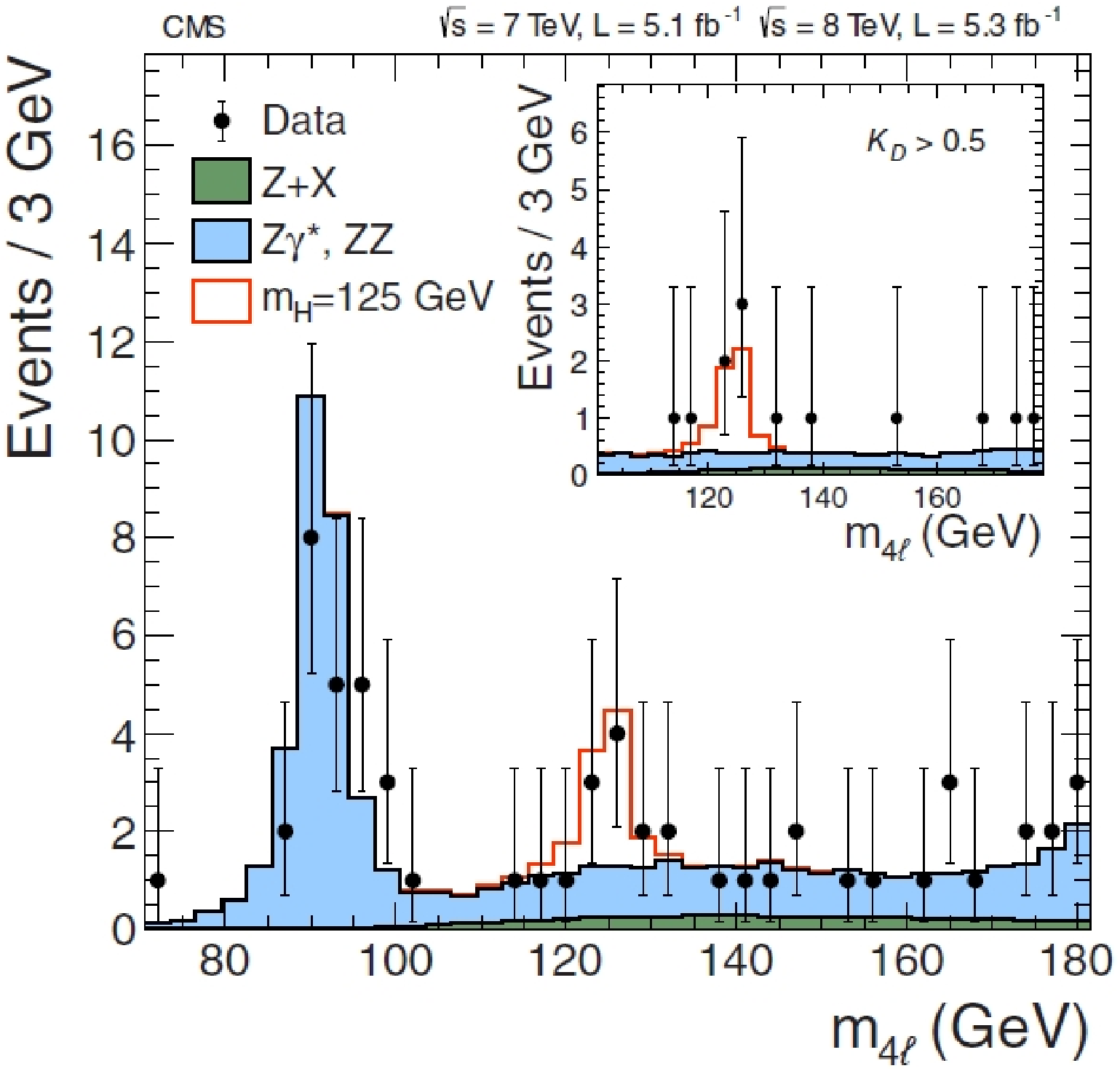}
\end{center}
\caption{\label{fig:cms_h2llll}Observation of the Higgs-like boson by CMS
  \cite{ref:cms_higgs_20120817} in the $\ell^+\ell^-\ell^+\ell^-$ invariant mass
  distribution at 125 GeV/$c^2$. The amplitude of the observed signal is close
  to the expectations of the Standard Model.}
\end{figure}

On 31 July the two experiments submitted papers of the discovery to Physics
Letters B, they were published 14 August \cite{ref:atlas_higgs_20120817,
  ref:cms_higgs_20120817}. Both papers are 15 pages long followed by 16 pages
of close to 3000 authors and both are dedicated to the memory of those
participants who could not live to see the result of the more than two decades
of construction work. Fig.~\ref{fig:cms_h2gg} shows the di-photon spectra
obtained by CMS in July 2012, after analyzing about a quarter of the data to
be collected in 2012. The 4-lepton spectra are quite similar with less
background and signal, for CMS it is
shown in Fig.~\ref{fig:cms_h2llll}.

\begin{figure}[t]
\begin{center}
\includegraphics[width=0.48\linewidth]{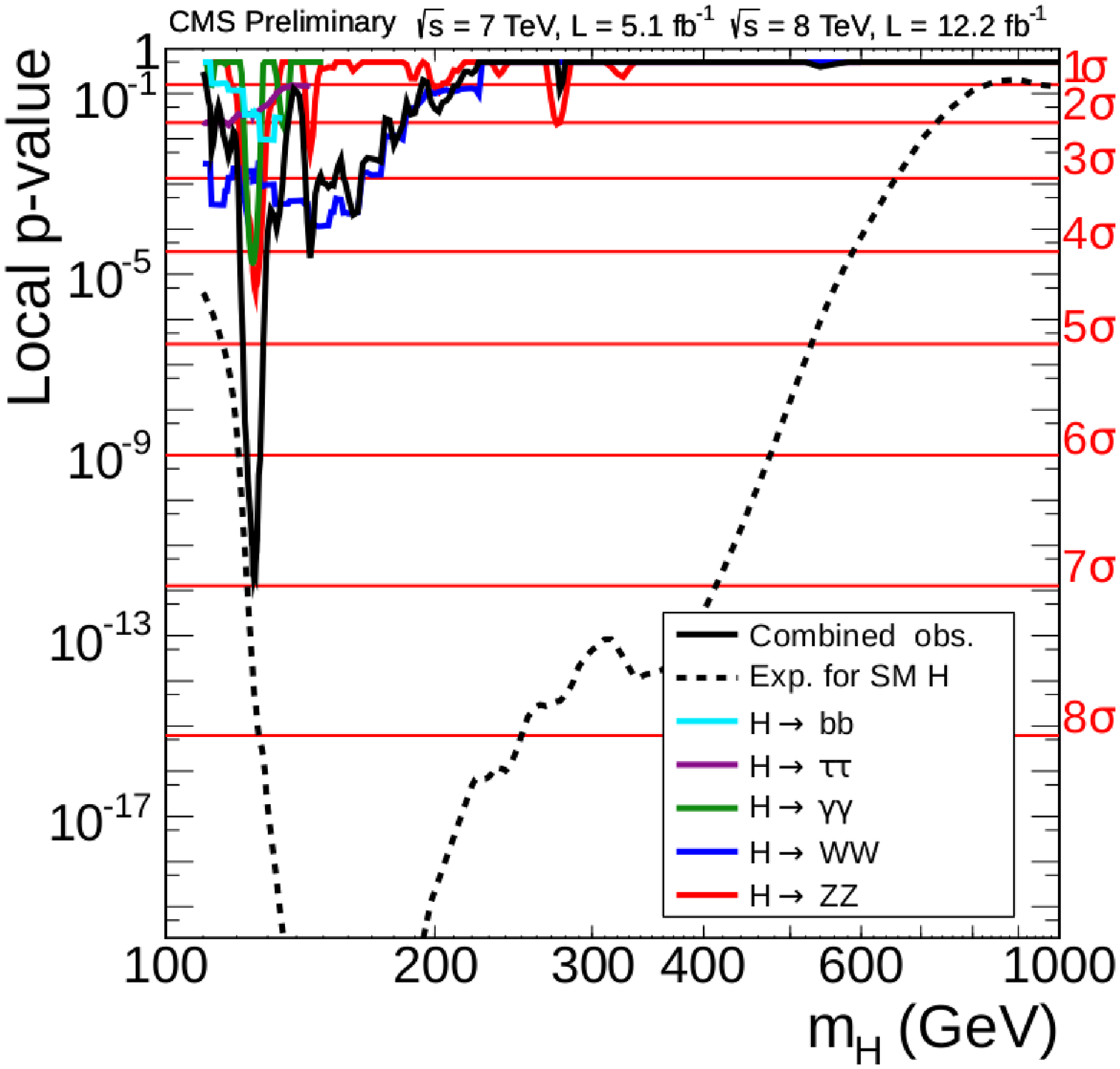}~
\includegraphics[width=0.48\linewidth]{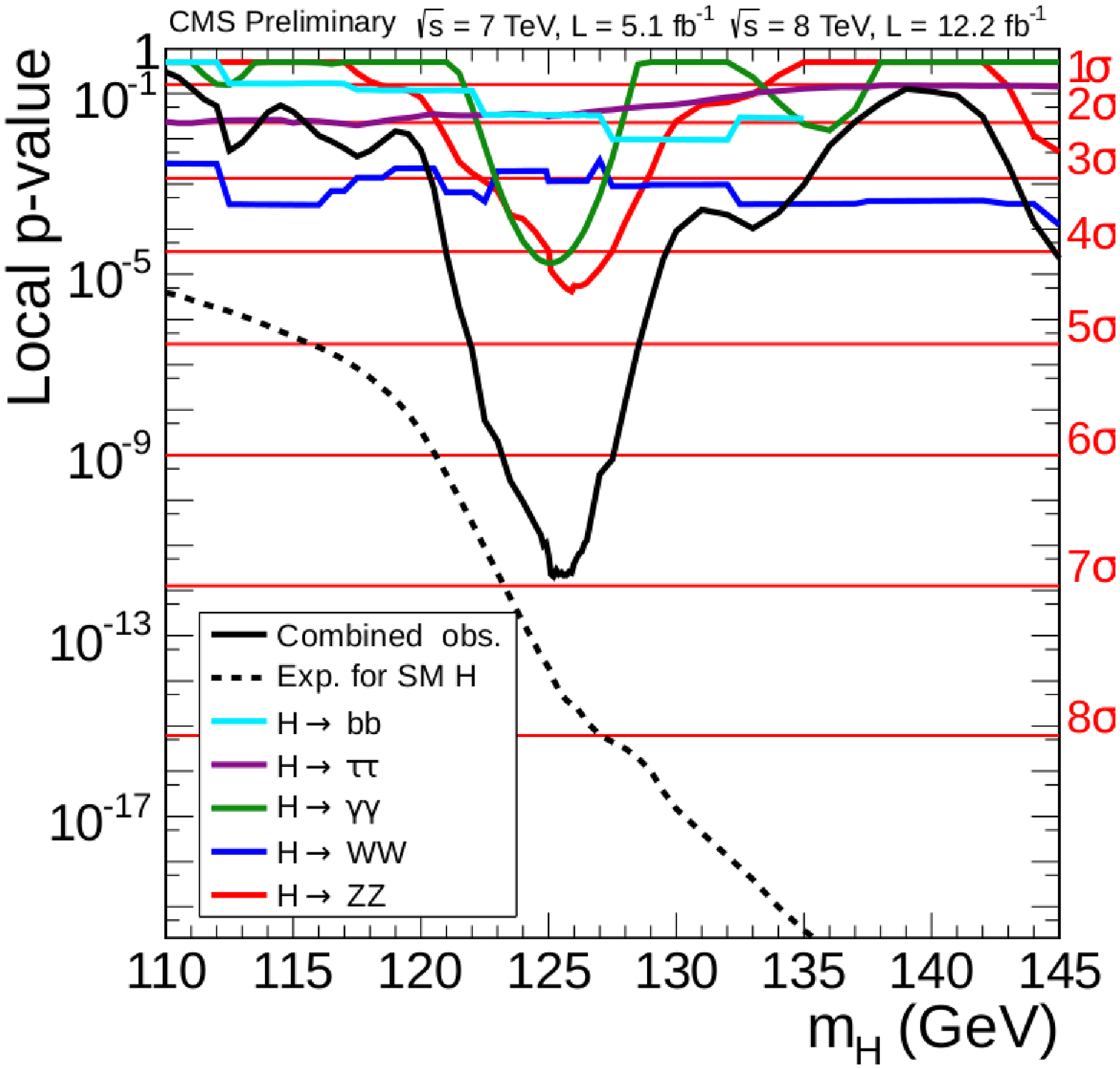}
\end{center}
\caption{\label{fig:cms_p-val}Observation of the Higgs-like boson by CMS
  \cite{ref:cms_HIG-13-005} in the invariant mass distribution of p-values in
  a wide range and closer to 125 GeV/$c^2$ as based on the data collected at
  LHC in 2011 and 2012. The amplitude of the observed signal is close to the
  expectations of the Standard Model.}
\end{figure}

What was really convincing of the observation was the distribution of the
p-values of the events selected in the various analyzed decay channels of the
hypothetical Higgs boson. For CMS data it is shown in Fig.~\ref{fig:cms_p-val}
for all available data: the significance is already as high as $6.9\sigma$.
It was a joke of statistics that in July 2012 adding together two decay
channels, H~\ra~$\gamma\gamma$ and H~\ra~$4\ell$ gave the same $5\sigma$
significance for both ATLAS and CMS whereas adding to it the WW channel
increased the significance to $6\sigma$ for ATLAS and reduced it to
$4.9\sigma$ for CMS.

\section{Is it the Higgs boson?}

Analyzing most of the data collected in 2012 led to the conclusion that all
observed properties of the newly discovered particle are within statistics
close to those predicted for the Higgs boson of the Standard Model. The fact
that it decays to two photons points to its having spin 0 or 2. The charged
lepton spectra bears the features of its having
$S=0^+$~\cite{ref:cms_2012_spin}. Its mass as determined by 
CMS~\cite{ref:cms_HIG-13-005} 
by the average of all decay channels is 
$<M_\mrm{X}> = \mrm{125.7 \pm 0.3 (stat) \pm 0.3 (syst)}$. The ATLAS result
is almost exactly the same: 
$\mrm{125.5 \pm 0.2 (stat) 
\left\{\begin{array}{l} +0.5\\ -0.6\end{array}\right\} (syst)}$. The
difference in the uncertainties are due to the facts that (i) ATLAS had more
signal-like data, but (ii) got more different masses in the two main
channels. The signal strengths of the new particle is also compatible with
that expected for the Standard Model Higgs boson: for CMS it is $0.80\pm0.14$
and for ATLAS $1.43\pm0.16$ (stat) $\pm0.14$ (syst). As a theoretician remarked
whenever ATLAS has an excess CMS comes up for everybody's annoyance with a
deficit, bringing the average close to the SM prediction. 

\begin{figure}[t]
\begin{center}
\includegraphics[width=0.33\linewidth]{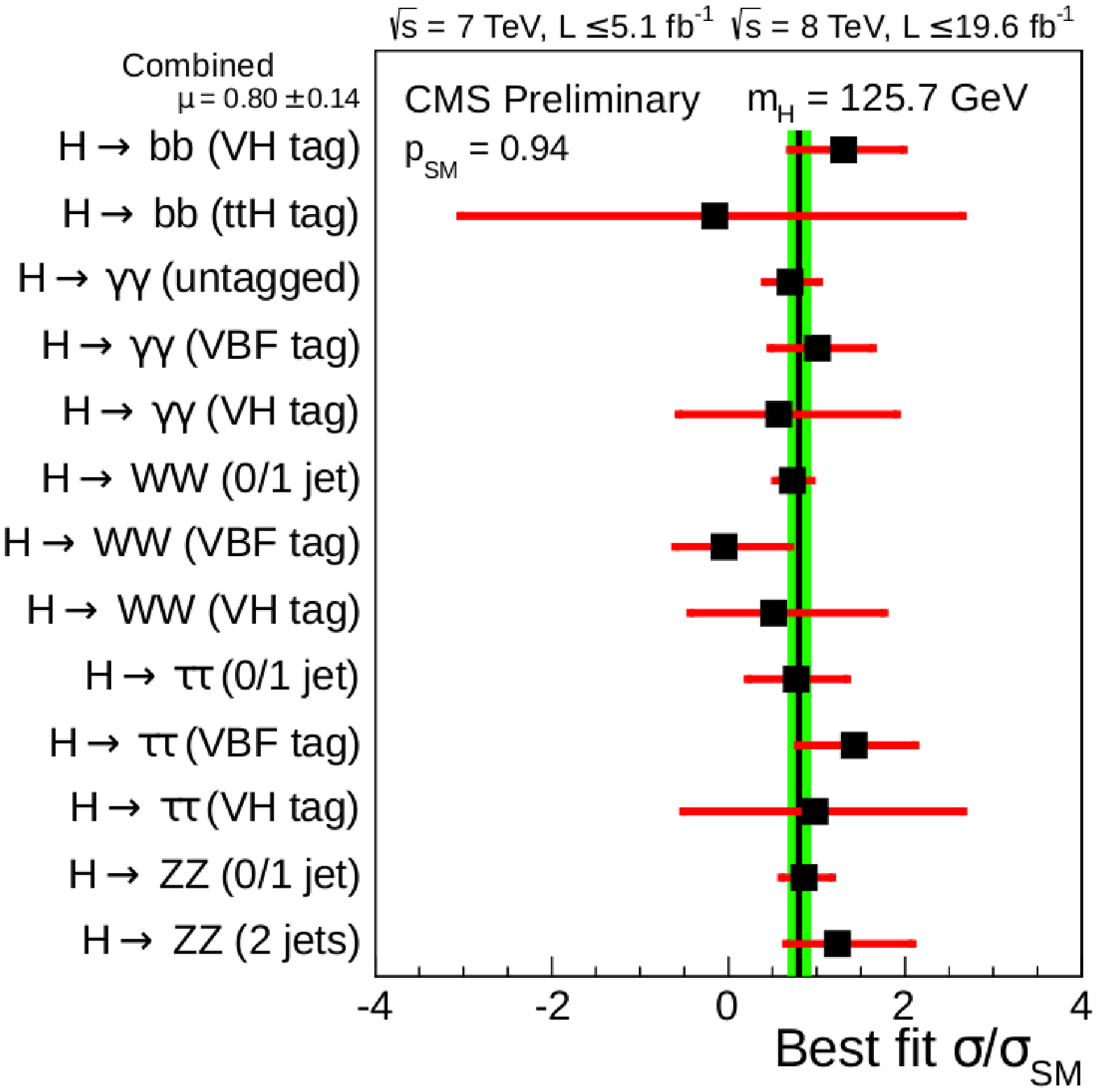}
\hspace*{-2mm}\includegraphics[width=0.33\linewidth]{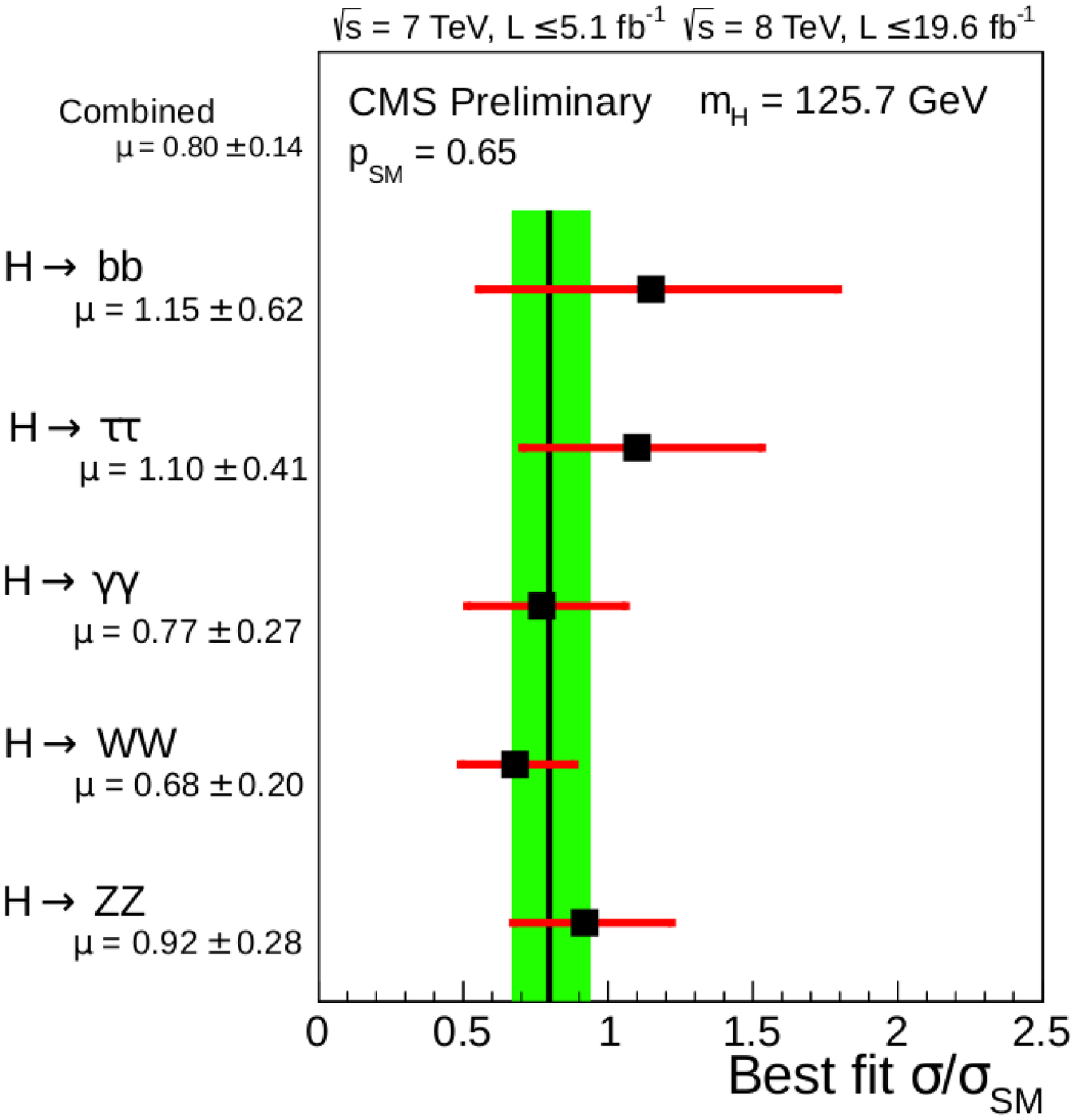}
\hspace*{-2mm}\includegraphics[width=0.33\linewidth]{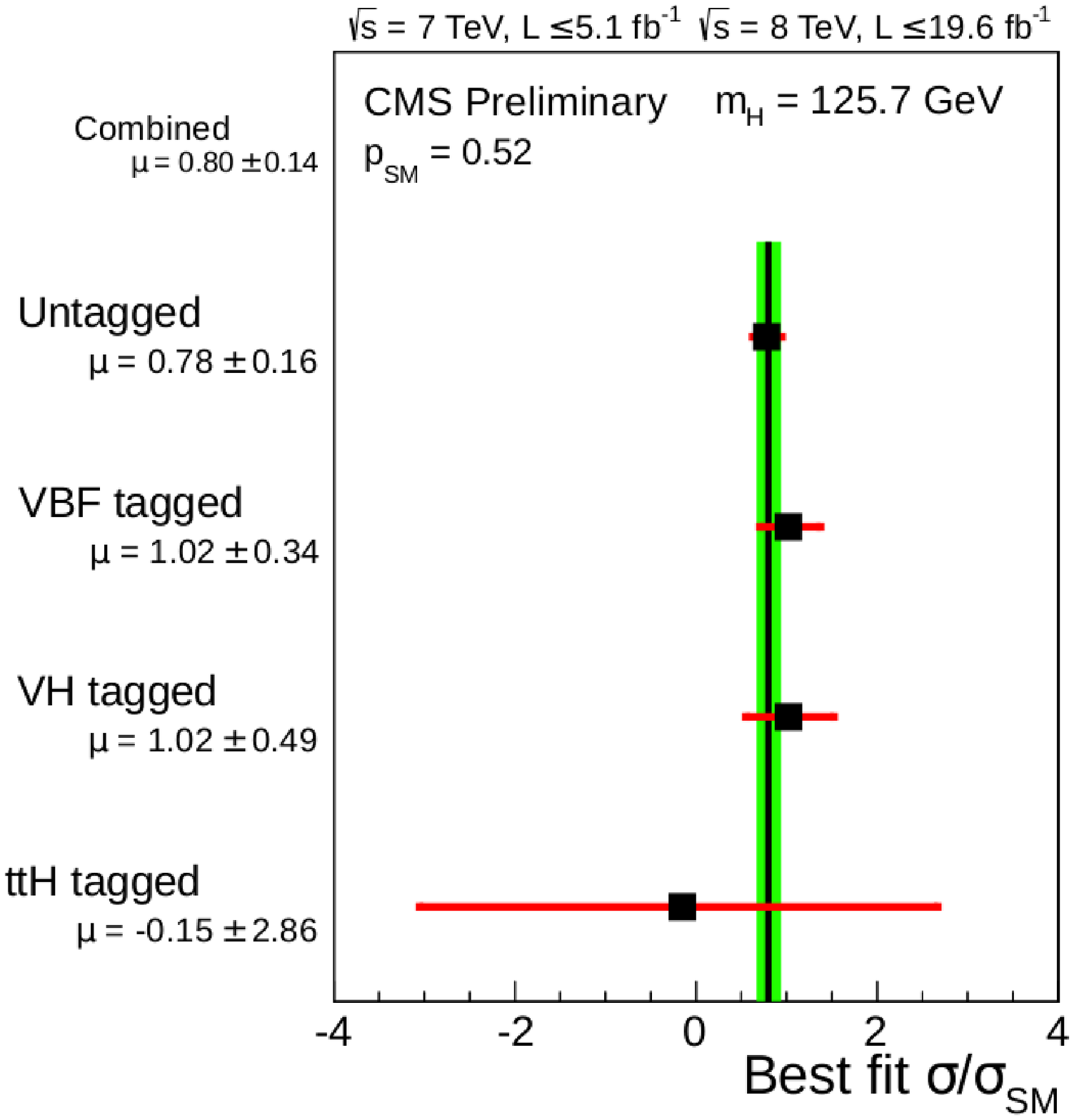}
\end{center}
\caption{\label{fig:cms_sig-str}Signal strengths of production and
  decay in various possible channels of the Higgs-like boson measured by CMS
  \cite{ref:cms_HIG-13-005} as compared to those predicted by the Standard 
  Model for the Higgs boson with a mass of 125 GeV/$c^2$.}
\end{figure}

The LHC experiments studied the cross sections of the processes connected to
the new particle. Fig.~\ref{fig:cms_sig-str} shows the signal strengths of
production and decay in various possible channels of the Higgs-like boson
measured by CMS \cite{ref:cms_HIG-13-005} as compared to those predicted by
the Standard Model for the Higgs boson with a mass of 125 GeV/$c^2$.  The
amplitudes of all observed signals are in agreement with the expectations of
the Standard Model.

Thus what we found is very likely the Standard Model Higgs boson. On one hand
this is a great success of particle physics. On the other hand this is
somewhat of a disappointment as the SM has theoretical shortcomings which need
new physics to resolve. Just to list a few of them: it cannot unite the
interactions at large energies, cannot account for the dark matter of the
Universe and cannot explain neutrino oscillations. There are many extensions
of the theory which should result in deviations from the Standard Model. All
those problems can be resolved e.g.\ by {\em supersymmetry}, but none of its
predicted phenomena could be found yet experimentally. The observables of the
Higgs boson should be sensitive to some of the features of new physics and
these studies will be the main job of ATLAS and CMS in the future, from 2015
when the LHC will restart with twice the energy and luminosity of 2012.

It is very interesting that the 126 GeV mass of the Higgs boson seems to be
exciting for theoreticians, there was even a special workshop
\cite{ref:whm126} organized to discuss this mass in 2013. The reason is that
$M_H=126$~GeV is at the border line of the stability of electroweak vacuum on
the plane of top mass against Higgs mass, see e.g.~\cite{ref:alekhin12}. At
the Madrid workshop the apparent fine tuning of the Standard Model compelled
some physicists to recall the anthropic principle.
 
\section{Acknowledgments}
The author is indebted to the organizers of the CCP-2013 conference for the
invitation with a substantial financial support. The results reported here are
due to the Higgs groups of LEP and of the CMS Collaboration. Tommaso Dorigo,
the chairperson of the CMS Statistics Committee has read the first draft of
the slides of this talk and improved it by very useful suggestions. This work
was supported by the Hungarian National Science Foundation OTKA via contracts
NK-81447 and K-103917.

\sloppy

\end{document}